\documentclass[%
reprint,
superscriptaddress,
 amsmath,amssymb,
 aps,
prb,
]{revtex4-2}

\usepackage{graphicx}
\usepackage{dcolumn}
\usepackage{ulem}
\usepackage{bm}
\usepackage[colorlinks=true,linkcolor=blue,citecolor=blue]{hyperref}
\renewcommand{\figurename}{\textbf{Figure}}
\usepackage{siunitx}
\usepackage{xcolor}

\begin{document}

\title{Berry phase manipulation in ultrathin SrRuO$_3$ films}

\author{Liang Wu}
 \email{lw590@physics.rutgers.edu, Present address: Kunming University of Science and Technology, Kunming, Yunnan, China. liang-wu@kust.edu.cn}
\affiliation{Department of Physics and Astronomy, Rutgers University, Piscataway, NJ 08854, USA}%
\author{Fangdi Wen}%
 \email{fw113@physics.rutgers.edu}
\affiliation{Department of Physics and Astronomy, Rutgers University, Piscataway, NJ 08854, USA}%
\author{Yixing Fu}
\affiliation{Department of Physics and Astronomy, Rutgers University, Piscataway, NJ 08854, USA}
\affiliation{Center for Materials Theory, Rutgers University, Piscataway, NJ 08854, USA}
\author{Justin H. Wilson}
\affiliation{Department of Physics and Astronomy, Rutgers University, Piscataway, NJ 08854, USA}
\affiliation{Center for Materials Theory, Rutgers University, Piscataway, NJ 08854, USA}
\author{Xiaoran Liu}
\affiliation{Department of Physics and Astronomy, Rutgers University, Piscataway, NJ 08854, USA}
\author{Yujun Zhang}
\affiliation{Institute for Solid State Physics, University of Tokyo, 5-1-5 Kashiwanoha, Chiba 277-8581, Japan}
\author{Denis M. Vasiukov}
\affiliation{Department of Physics and Astronomy, Rutgers University, Piscataway, NJ 08854, USA}
\author{Mikhail S. Kareev}
\affiliation{Department of Physics and Astronomy, Rutgers University, Piscataway, NJ 08854, USA}
\author{J. H. Pixley}
\affiliation{Department of Physics and Astronomy, Rutgers University, Piscataway, NJ 08854, USA}
\affiliation{Center for Materials Theory, Rutgers University, Piscataway, NJ 08854, USA}
\author{Jak Chakhalian}
\affiliation{Department of Physics and Astronomy, Rutgers University, Piscataway, NJ 08854, USA}

\date{\today}

\begin{abstract}

 Berry phase is a powerful concept that unravels the non-trivial role of topology in phenomena observed in chiral magnetic materials and structures. 
 A celebrated example is the anomalous Hall effect (AHE) driven by the non-vanishing Berry phase in momentum space. As the AHE is dependent on details of the band structure near the Fermi edge, the Berry phase and AHE can be altered in thin films whose chemical potential is tunable by dimensionality and disorder. 
 Here, we demonstrate how Berry phase in ultrathin SrRuO$_3$ films provides a comprehensive explanation for the non-trivial Hall response which is conventionally attributed to the topological Hall effect (THE). 
 Particularly, the Berry phase contribution to this effect can be altered, enhanced, and even change signs in response to the number of layers, temperature, and importantly, disorder. 
 By comparing the effects of disorder theoretically on a skyrmion model and a spin-orbit coupled model, we show that disorder suppresses the THE while it can enhance  the AHE. 
 The experiments on more disordered samples  confirm this interpretation, and  proposed  multi-channel analysis  judiciously explains the observed THE-like feature.

\end{abstract}

\maketitle

Recently, the search for topologically non-trivial modalities has become a dominant driver in condensed matter physics \cite{bruno,1,2,3,4,5,6}. A metallic magnet entwined with a non-collinear spin texture like skyrmions, chiral domain walls, or helical order can demonstrate interesting phenomena due to the interaction of conduction carriers with the localized spins \cite{5,6,7,8,manchon}. 
Such non-trivial magneto-transport response stems from the emergent electromagnetic fields (EEMFs) linked to finite Berry phase accumulation. 
Experimentally, the challenge is to devise clear signatures of the EEMFs linked to the topologically spin texture with  a non-zero winding number or skyrmions. 
In fact, such skyrmionic contribution can be revealed as the extra  features in the transverse  Hall resistivity $\rho_{xy}$ arising from a fictitious Lorentz force \cite{4,9}, which is collectively known as the topological Hall effect or THE. 
Because of the simplicity of the detection method, THE has become a popular tool to search for the presence of skyrmionic matter in bulk crystals, thin films, and hetero-junctions \cite{10,11,12,13,14,15,16,17,18,19,20,21,22,23}. 

However, the criterion to demonstrate the THE is to observe additional bumps and dips that contribute to the (non-quantized) Hall resistance as a function of magnetic field. These features can only be definitively attributed to the THE under the \textit{assumption of homogeneous materials with a single conduction channel} \cite{24,25,26,27, yang, shen}. 
Moreover, the unavoidable effects of disorder in ultra-thin layers, can drastically alter the Berry phase and the resulting AHE, and yet are typically not considered.
In this work, we test the validity of the THE interpretation experimentally and theoretically by focusing on a prototypical ferromagnetic metal SrRuO$_3$ (SRO), with perovskite structure (Pnma) and a bulk Curie temperature $T_c \approx \SI{150}{K}$ \cite{28,29,30}. For this work, samples are grown in ultrathin form to amplify  the effects of confinement, as well as magnetic and structural inhomogeneity on the Berry phase and AHE. 
Importantly, to verify the THE scenario we use two parameters, disorder and film thickness on the observed AHE.
By including these two factors when analyzing the experimental data with a two-channel AHE, we successfully reproduce both the overall transverse Hall effect and the universal  scaling  behavior between AHE conductivity  $ \sigma_{\mathrm{AHE}} $ and longitudinal conductivity  $ \sigma_{xx} $ without invoking the THE.
Theoretically, we compare the effects of disorder on two opposing models of SRO: (1) a model of THE with skyrmions \cite{20,21,22,23} or (2) a  model of spin-orbit coupling induced fluctuating Berry phase \cite{31}. As a result, only the latter model is able to  capture the experimentally observed  tunability of the AHE with film-thickness, disorder, and temperature.

 \begin{figure}[htp]
	\includegraphics[width=\columnwidth]{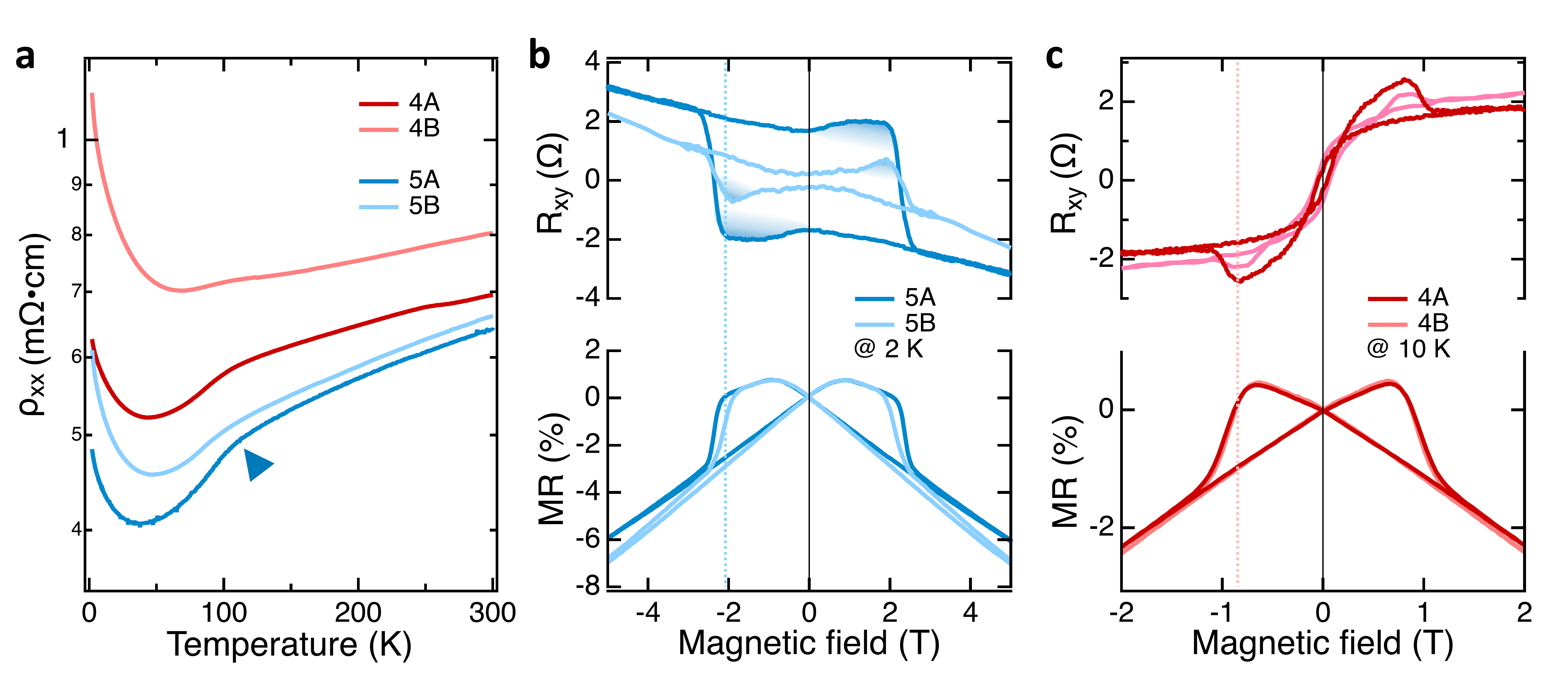}
	\caption{\textbf{Longitudinal and transverse transport results of ultrathin SRO films.} \textbf{a}, temperature dependence of resistivity $\rho_{xx}$, red and blue curves indicate the samples with as-grown (A-series) and air-exposed (B-series) conditions, respectively. \textbf{b-c}, MR and Hall measurements of 5 u.c.\ and 4 u.c.\ samples at 2 and 10 K, respectively. Note, the shaded areas are attributed to the THE-like Hall contribution.}
	\label{Fig1}
\end{figure}

\begin{figure}
	\includegraphics[width=\columnwidth]{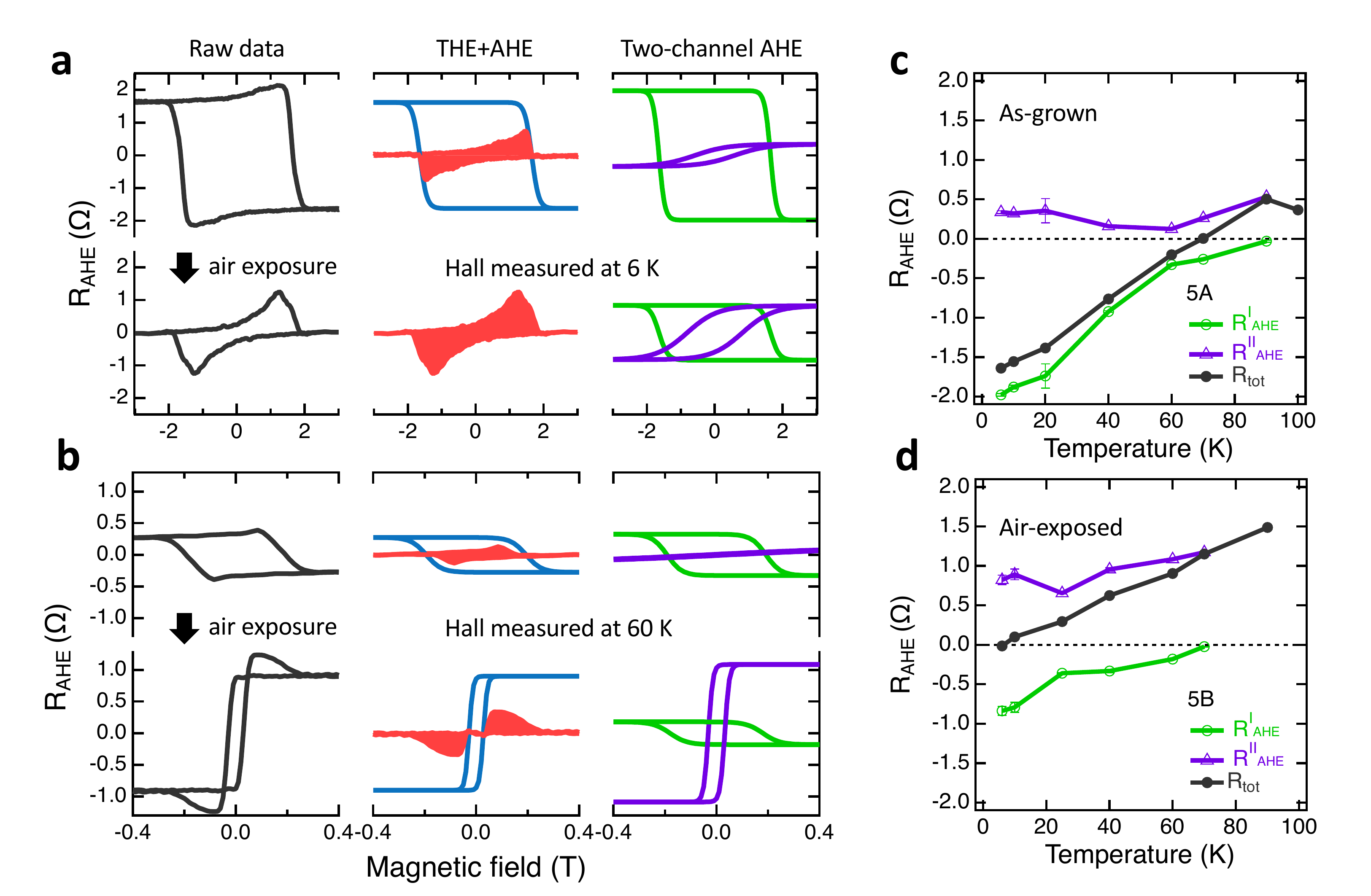}
	\caption{\textbf{Disentanglement of Hall signal in 5 u.c.\ SRO with THE and two-channel AHE scenarios, respectively.} \textbf{a}, Upper panel shows the experimental Hall data, fitting results with THE and two-channel AHE scenarios at 6 K of as-grown sample 5A. Lower panel is the identical treatment of Hall data for the same sample after air exposure (5B). The red shaded area and blue curves indicate the $R_{\mathrm{THE}}$ and $R_{\mathrm{AHE}}$, respectively (middle panel). The green and purple curves indicate the $R^-_{\mathrm{AHE}}$ and $R^+_{\mathrm{AHE}}$, respectively (right panel). \textbf{b}, Same Hall data treatment for measuring temperature of 60 K, more temperature-dependent Hall data can be found in Ref. \cite{32}. \textbf{c-d}, Temperature-dependent total and two-channel AHE resistance.}
	\label{Fig2}
\end{figure}

To investigate the temperature dependence of the AHE as a function of confinement, we have acquired a series of temperature-dependent resistivity curves (R-T) shown in Fig.~\ref{Fig1}a. In addition to the pristine samples (labeled as series A samples of 4 u.c.\ and 5 u.c., 4A and 5A respectively), we have developed a protocol to introduce controlled disorder by time-exposing the as-grown samples to ambient conditions (labeled as series B samples, 4B and 5B). (Ref. \cite{32})
As seen, all films display a common feature: a metallic state at high temperature range and a small kink at around 100 K. 
This is the paramagnetic-to-ferromagnetic phase transition, which is lower than that of the bulk SRO caused by  the ultrathin nature of  the samples\cite{20,21,33}. 
In addition, a characteristic upturn appears below 40 K, which is due to  disorder-induced Anderson weak localization in the ultrathin limit \cite{anderson, 34,35,36}. 
Further, as seen in Fig.~\ref{Fig1}(b,c) for both 4B and 5B samples, the R-T curves still retain the characteristic shape of the as-grown samples 4A and 5A, albeit with larger resistivity and steeper upturn at the low temperature, indicating increased disorder \cite{anderson, 34,35,36}.
Importantly, the magnetoresistance (MR) measurements show only a negligible difference between the A and B series samples (see Fig.~\ref{Fig1}b, c) implying that disorder  barely impacts ferromagnetism in SRO.
With  this understanding, we conduct temperature-dependent Hall measurements to investigate the thickness- and disorder-dependent AHE at different temperatures.
Figures~\ref{Fig1}(b,c) show  the representative low-temperature Hall resistance data.
Following convention, the magnitude of anomalous Hall resistance is extrapolated from the high-field linear part of the data, whose sign defines the sign of AHE, namely, the AHE in 4A  is refereed to positive whereas in  5A it is  negative. 
This attribution is also consistent with the previous reports \cite{20,21,37}. 
However, Fig.~\ref{Fig1}(b,c) reveals an unexpected result that the Hall data undergo a strong change for the B-series samples. 

To  elucidate this,   we apply the data analysis  which is conventionally used to extract THE and separate various  contributions to  the  Hall effect in SRO. 
Under the assumption of  the  presence of  skyrmions, the Hall resistance can be decomposed as $ R_{xy}=R_{\mathrm{OHE}}+R_{\mathrm{AHE}}+R_{\mathrm{THE}}$ \cite{20,21,37}, where $R_{\mathrm{OHE}}$ stands for ordinary Hall effect (OHE). Alternatively, the THE-like Hall resistance can also be decomposed  as $ R_{xy}=R_{\mathrm{OHE}}+R^+_{\mathrm{AHE}}+R^-_{\mathrm{AHE}} $, 
where the last two terms denote positive and negative contributions to the two-channel AHE \cite{24,26}.
In the following, all OHE has been subtracted by fitting the high field linear slope.
Given that thin-film, complex oxide perovskites often exhibit intrinsic propensity for defects and layer non-uniformity during the step-flow growth \cite{choi}, we demonstrate that the minimal two-channel AHE without THE successfully captures all the AHE features  across the entire temperature range.
 First, we note that a thickness variation of at least 1 u.c.\ exists in a real film since step edges on the surface of an SRO film cannot ideally replicate those of the substrate  after the growth (Ref. \cite{32}).  
  Experimentally,  this thickness variation in SRO was recently observed by magnetic force microscopy \cite{21,37,38}. In addition, the two-step transitions found in magnetic hysteresis loops \cite{33,39} and  MR shown in Fig.~\ref{Fig1}(b) lend strong support for the thickness variation of the nominal SRO thickness.
  This implies that films nominally grown as 5 u.c. will in fact vary from 4 u.c. to 6 u.c. across the sample.
   This intrinsic thickness non-uniformity, whose effect can often be neglected in thick films, can markedly alter the transport properties upon approaching the ultra-thin limit.

As seen in Fig.~\ref{Fig1}, the sign of overall AHE is \textit{opposite} for the 4A film to that of the 5A below the crossover temperature around 90K \cite{20,21,37}. 
Given the intrinsic variation of the thickness, we are able to reproduce the total AHE of the nominal 5 u.c. sample  as the sum of the positive (4 u.c.) and negative AHEs (5 u.c.) originating  from the thickness variation throughout the SRO film,  \textit{without} the `superficial' THE feature. 
Meanwhile, we also have contributions from $\gtrsim 6$ u.c.\ regions, we find films grown at 6 u.c.\ and above only contribute a negative AHE and thus these regions are included in the 5 u.c.\ signal $R^-_{\mathrm{AHE}}$ \cite{20,21,22,23}.
As the temperature is raised, the sign of the negative AHE switches and the THE-like Hall bump  disappears as now both AHE channels become positive, entirely consistent with the two-channel scenario.

 Next, we quantify the individual AHE contributions within the two-channel model in 5 u.c.\ SRO, while 4 u.c.\ SRO data can be found in Ref. \cite{32}. 
Separating the two channels $R^{\mathrm{tot}}_{\mathrm{AHE}}= R^\mathrm{I}_{\mathrm{AHE}}\tanh(\omega_\mathrm{I}(H-H^\mathrm{I}_\mathrm{c})) + R^{\mathrm{II}}_{\mathrm{AHE}}\tanh(\omega_{\mathrm{II}}(H-H^{\mathrm{II}}_\mathrm{c}))$, where $H_\mathrm{c}$ denotes the coercive field and $\omega$ is the slope of $R_{AHE}^{tot}$ at $H_c$ for each channel. 
In contrast, we also complete the data analysis based on the THE scenario, $R_{xy}=R_{\mathrm{AHE}}+R_{\mathrm{THE}}$. Figure~\ref{Fig2}a shows the experimental Hall data and results of the fitting by both models for sample 5A (top panel) and 5B (bottom panel) at \SI{6}{K} while Fig.~\ref{Fig2}b represents the same data treatment at \SI{60}{K}. 
 More detailed temperature- and thickness-dependent analysis can be found in Ref. \cite{32}.
 Direct inspection of Fig.~\ref{Fig2} confirms that the fitting curves by the two-channel AHE model are indeed in excellent agreement with the experimental data with no  additional THE-like contribution. 
 In addition, the temperature-dependent total and the two-channel AHE signals are summarized in Fig.~\ref{Fig2}(c,d).
 
We note that the increased disorder in air-exposed samples not  only can increase the positive Hall contribution, but also it  can reverse the sign of the overall Hall signal. 
This remarkable effect can be  understood by considering the concentration of Berry curvature in momentum space and how disorder (as well as thickness and temperature) ``smear'' out different contributions.
The overall result  depends on the band structure near the Fermi edge \cite{40}. 
The mechanism of disorder-, thickness-, and temperature-dependent AHE in our sample is  discussed in detail in the  theory part.
\begin{figure}[htp]
	\includegraphics[width=\columnwidth]{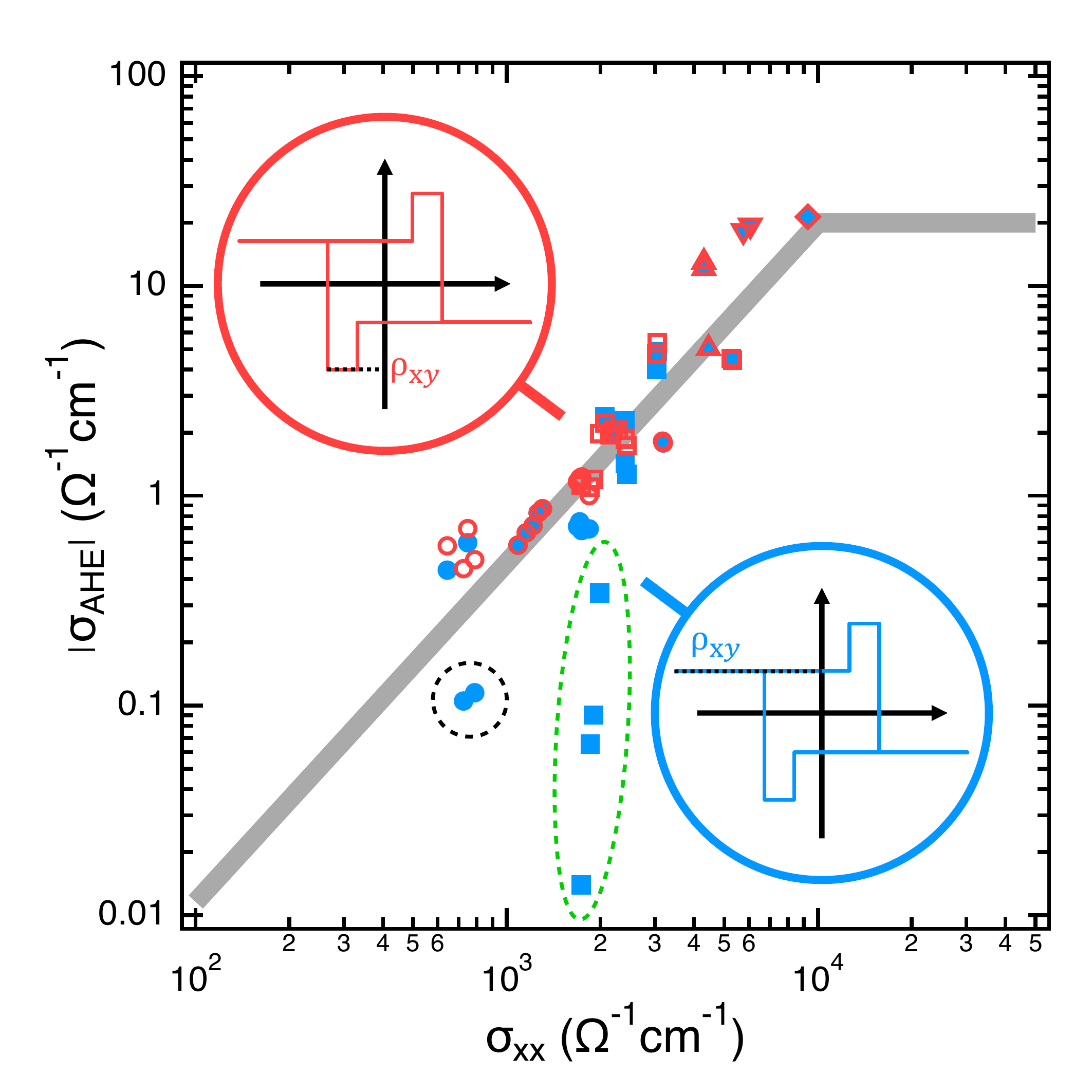}
	\caption{\textbf{Comparison of the scaling relation.}   Blue filled and red open symbols are scaling relation derived in THE scenario and two-channel AHE scenario, respectively. The black and green dash circles indicate the large deviation from the universal scaling relation of BaTiO$_3$-capped 4 u.c.\ SRO \cite{21} and 5B, where large putative THE-like Hall signal appears simultaneously. The circle, square, up-pointing triangle, down-pointing triangle, and diamond symbols indicate 4, 5, 6, and 7 u.c. samples, respectively.}
	\label{Fig3}
\end{figure}

To further verify the two-channel AHE model, we apply the scaling relation between the magnitude of anomalous Hall conductivity $| \sigma_{\mathrm{AHE}} |$  and the longitudinal conductivity $\sigma_{xx}$. This universal scaling relation divides materials that exhibit the AHE into three regimes \cite{25,41,42,43}. The ultra-thin SRO with $\sigma_{xx}\leq$ $10^4 \ \Omega^{-1}\textnormal{cm}^{-1}$ belongs to the bad metal regime, where disorder smears the contribution from any intrinsic Berry phase driven AHE, resulting in the scaling relation  $\sigma_{\mathrm{AHE}}\sim(\sigma_{xx})^{1.6}$. Here, we note that in the THE scenario, the extra putative THE contribution to the Hall measurements would vanish at a sufficiently high magnetic field due to magnetic moments aligning with the magnetic field, destroying any non-trivial spin textures \cite{20,21,37}. Therefore, the pure AHE contribution extracted from the high magnetic field regime should follow the universal scaling relation.
However, if the purported THE signal is induced  by a two-channel AHE with opposite signs, the overall AHE extracted from the high magnetic field regime is merely a result of a (partially) canceled superposition of two-channel AHE,  $|R_{\mathrm{overall}}|=||R^{+}_{AHE}|-|R^{-}_{AHE}|| $. Moreover, since each AHE channel should follow the universal scaling relation, the destructive overall AHE signal would deviate from the universal scaling relation and appear lower than the expected  relation of $\sigma_{\mathrm{AHE}}\sim(\sigma_{xx})^{1.6}$. If each channel obeys this scaling but contributes oppositely to transport, $R_{\mathrm{overall}}$ could show deviates from the universal scaling. To alleviate this deviation, a constructive superposition of $ |R_{\mathrm{overall}}|=|R^{+}_{AHE}|+|R^{-}_{AHE}| $ is used to represent the two-channel AHE scenario with opposite signs, which is equivalent to the two-channel AHE scenario with same sign (Ref. \cite{32}).
  Fig. \ref{Fig3} shows the experimental results based on both THE  and two-channel AHE scenarios along with the summary of previously published SRO data \cite{20,21}. The samples with large putative THE-like Hall signal diverge from the universal scaling, whereas our scaling data derived from the two-channel AHE scenario show \textit{excellent consistency with the universal scaling relation}.

Finally, we discuss the physical mechanism of the observed thickness, disorder, and temperature dependence on the AHE sign reversal in ultra-thin SRO films Figure~\ref{Fig1} shows the results of  numerically exact calculations on a 
minimal lattice model of SRO~\cite{31} that produces an AHE due to a spin-orbit coupling induced Berry curvature. In bulk three-dimensional models, temperature can induce a sign reversal in the AHE \cite{31}. Here, we demonstrate that in disordered thin-film samples varying the thickness and/or disorder can induce a reversal of sign and an enhancement of the AHE. Additionally, we  analyze a separate theoretical model purported to capture  the THE in SRO and show the effect of disorder on the THE is incompatible with the experimental data (Ref. \cite{32}).

We apply disorder to thin films of a spin-polarized tight-binding model with $d_{xz}$ and $d_{yz}$ orbitals (originating from the $t_{2g}$ orbitals of the cubic perovskite structure) from Ref.~\onlinecite{31} known to host an AHE. The model is on a square lattice (in the 1 u.c.\ limit) with spin-orbit coupling, with a finite magnetization, and with orbitals that induce nearest and next-nearest neighbor hopping (with strength $t_1$ and $t_2$ respectively) on each individual u.c. while each u.c. is tunnel-coupled (with strength $t_1$).
We consider slab sizes of $L_x=L_y=L=233$ and vary the number of u.c.\ $L_z = 1 - 7$ with open boundary conditions in the $z$-direction and periodic in the $x$- and $y$-direction.
While the features we see appear generic over different parameters and types of the disorder (Ref. \cite{32}), we focus here on on-site potential disorder with strength $W$.
Using the kernel polynomial method \cite{44}, we  compute the full conductivity tensor~\cite{45} and hence the resistivity.
The exact model, details of the calculation, and a broader consideration of the parameters and finite size can be found in Ref. \cite{32}.

\begin{figure}
	\includegraphics[width=\columnwidth]{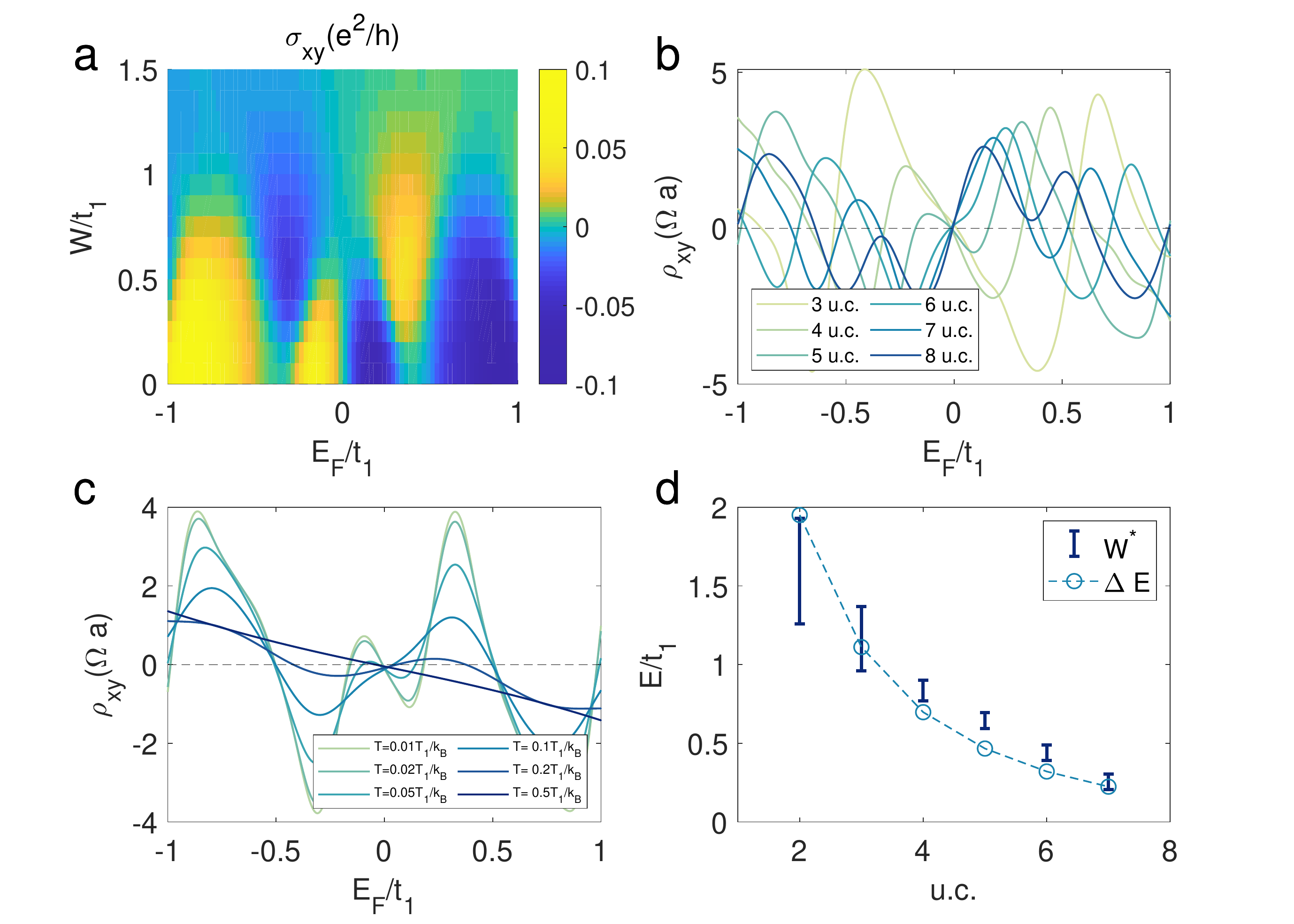}
	\caption{\textbf{a}, Anomalous Hall conductivity $\sigma_{\mathrm{AHE}}$ vs disorder strength $W$ and Fermi energy $E_F$. \textbf{b-c}, Anomalous Hall resistivity $\rho_{\mathrm{AHE}}$ versus $E_F$ for different \textbf{b} number of u.c.\ and \textbf{c}, temperatures.
		Notice that for each case there is a particular value of the changing variable where the AHE changes sign. In the case of disorder strength, we call this value $W^*$.
		For \textbf{a} and \textbf{b} the temperature is held at $T=0.025 t_1/k_B $, \textbf{b} and \textbf{c} hold the disorder constant at $W=0.5 t_1$ , and \textbf{a} and \textbf{c} are for 5 u.c..
		In \textbf{d} we show the relation between $W^*$ and number of u.c.\ as well as the distance between the Van Hove peaks nearest to $E_F=0$ in the disorder free limit for $T=0.025 t_1/k_B$.
		These quantities are correlated due to how they are related to the Berry curvature distribution within the bands. }
	\label{fig:theory}
\end{figure}

The numerical results are presented in Fig.~\ref{fig:theory} for the Hall resistivity $\rho_{xy}$. In the theory section, the only contribution for $\rho_{xy}$ is $\rho_{\mathrm{AHE}}$.
At a particular disorder strength that we label $W^*$, the sign of the AHE changes.
We precisely define this as the point at $E_F=0$ when the slope of $\rho_{xy}$ changes signs. Our results clearly demonstrate that the sign of $\rho_{xy}$ changes due to varying the strength of disorder (Fig.~\ref{fig:theory}a), the number of layers (Fig.~\ref{fig:theory}b), and the temperature (Fig.~\ref{fig:theory}c).
In Fig.~\ref{fig:theory}d we fix the temperature and find $W^*$ as a function of number of u.c..
If we contrast the change in the AHE between disorder and temperature, we notice that increasing disorder is providing an enhancement of $\rho_{xy}$ near $E_F\approx 0.5t_1$ in Fig.~\ref{fig:theory}a whereas increasing temperature is suppressing this feature Fig.~\ref{fig:theory}c.
In the experiment, thinner films have larger disorder, and thus the experiment is taking a cut across the $W^*$ boundary in Fig.~\ref{fig:theory}d. As we cannot disentangle these two effects, the theoretical calculations demonstrate that the experimental results exhibit a sign 
reversal in the AHE due to varying both the number of layers and the strength of disorder. 
Importantly, this also demonstrates that disorder can enhance the AHE signal consistent with our experimental observations in Fig.~\ref{Fig2}. In the 
Supplemental Material \cite{32},  we show this cannot be accounted for using the skyrmion model for the THE where disorder leads to a featureless AHE.

In this few u.c. model, the density of states (Ref. \cite{32}) shows that each u.c. creates a pair of Van Hove peaks, with an offset due to hopping in the $z$-direction.
Using the Van Hove peaks as a guide, the distance between neighboring peaks near zero energy, denoted as $\Delta E$, reduces with increased number of u.c.\ and
Fig.~\ref{fig:theory}d shows that this trend is correlated with  $W^*$.
As the number of u.c.\ increases the sign reversal in $\rho_{xy}$ occurs  when the disorder strength smears the two van Hove peaks closest to $E_F=0$ into one.
This is a signal of the general phenomena: due to the nonuniform distribution of Berry curvature in the bands, the inclusion of terms which sample states with different Berry curvature (e.g.\ temperature, disorder, and thickness) can significantly alter the AHE, even changing its sign and enhancing its value.  

In conclusion, the presented results unveil the unexpected thickness-dependence in the AHE signal in the ultra-thin limit. 
Moreover, the long-overlooked disorder effect can drastically modify the AHE signal, to  the point of reversing  its sign.
Such extreme sensitivity stems from the effects of disorder on the intrinsic Berry phase contribution of the AHE as confirmed by numerically exact calculations with the kernel polynomial method on a model that hosts the AHE. 
In addition, our findings provide  experimental evidence for the superficial nature of the THE attribution to topologically  protected spin texture  and instead lend  strong support to the two-channel AHE in SRO. 
This is also supported theoretically by the study of disorder effect on the putative THE, which is found to be contradictory to the experimental data.
The proposed multi-channel magneto-transport framework can be readily  extended to many other ultra-thin chiral magnetic metals with  spin-order where disorder, and the effects of surface and interface are critically important.\medskip

\acknowledgements{
We deeply acknowledge fruitful and insightful discussions with Weida Wu, Daniel I. Khomskii,  and X. Renshaw Wang. This work was supported by the Gordon and Betty Moore Foundations EPiQS initiative through Grant No. GBMF4534 and
NSF CAREER Grant
No. DMR-1941569 (Y.F. and J.H.P.).
J.H.W. and J.H.P. acknowledge the Aspen Center for Physics where some of this work was completed, which is supported by National Science Foundation grant PHY-1607611.}


\medskip

\pagebreak

\end{document}